\documentclass[%
 aip,
 jmp,
amsmath,amssymb,
preprint,%
]{revtex4-1}

\usepackage{graphicx}% Include figure files
\usepackage{dcolumn}% Align table columns on decimal point
\usepackage{bm}% bold math
\usepackage[utf8]{inputenc}
\usepackage[T1]{fontenc}
\usepackage{mathptmx}
\usepackage{color}

\newcommand*{\ampd}[2]{
    t^{#1}(#2)
}
\newcommand*{\rhs}[1]{
    \xi^{#1}
}
\newcommand*{\lhs}[1]{
    \eta^{#1}
}
\newcommand*{\muld}[2]{
    {\bar{t}}^{#1}(#2)
}

\newcommand*{\rightv}[1]{
    R_{#1}
}
\newcommand*{\leftv}[1]{
    L_{#1}
}
\newcommand*{\rsp}[2]{\langle\!\langle #1; #2 \rangle\!\rangle} % needed for the appendix

\newcommand*{\cch}[1]{\textcolor{blue}{#1}}

\begin{document}

%\preprint{MCD-CC}

\title[MCD-CC]{Magnetic circular dichroism spectra from  resonant and
damped coupled cluster
%singles and doubles
response theory}
% Force line breaks with \\

\author{R. Faber}
\affiliation{DTU Chemistry, Technical University of Denmark, Kemitorvet Bldg 207, DK-2800 Kongens Lyngby, Denmark}
\author{S. Ghidinelli}%  
\affiliation{Department of Molecular and Translational Medicine, Universit{\`a} degli Studi di Brescia, Viale Europa 11, I-25123 Brescia, Italy.}
\author{C. H{\"a}ttig}
\affiliation{Arbeitsgruppe Quantenchemie, Ruhr-Universit{\"a}t Bochum, D-44780, Germany}
\author{S. Coriani}
\thanks{Author to whom correspondance should be addressed: soco@kemi.dtu.dk}
\affiliation{DTU Chemistry, Technical University of Denmark, Kemitorvet Bldg 207, DK-2800 Kongens Lyngby, Denmark}
 
\date{\today}% 

\begin{abstract}
A computational expression for the Faraday ${\mathcal{A}}$ {term}
of magnetic circular dichroism (MCD) is derived within coupled cluster response theory and alternative computational expressions for the ${\mathcal{B}}$ term are discussed. 
Moreover, an approach to compute the (temperature-independent) MCD ellipticity in the context of coupled cluster damped response is presented, and its equivalence with the stick-spectrum approach in the limit of infinite lifetimes is demonstrated. The damped response approach has advantages for molecular systems or spectral ranges with a high density of states. 
Illustrative results are reported at the coupled cluster singles and doubles level and compared to time-dependent density functional theory results.
\end{abstract}

\maketitle

\section{\label{sec:Intro}Introduction}

In magnetic circular dichroism (MCD) spectroscopy, the sample is probed with
circularly polarized light in presence of a relatively strong magnetic field oriented parallel to the direction of propagation of the light beam.
The external magnetic field induces a differential absorption of the right- and left- circularly polarized light.~\cite{MasonBook}
MCD can provide insight to the geometric, electronic, and magnetic properties of chemical systems.
The applied magnetic field couples to the (spin and/or orbital) angular momentum, lifting the degeneracies among {ground and} excited states (by Zeeman splitting), and giving rise to additional spectroscopic features compared to the zero-field case.
%
% The perturbation of the electronic states by the magnetic field allows to probe the nature of the state(s) involved in the electronic transition. This is the primary value of MCD. 
% It can assist in the interpretation of electronic transitions and can provide useful ground and/or excited state magnetic properties, which in some cases are difficult or impossible to obtain by any other means. 
% When combined with conventional absorption spectroscopy, MCD adds another experimental dimension: 
%
Since the MCD spectral features are signed and depend upon molecular magnetic moments in electronic states and the direction of the field, MCD yields additional information when combined with
conventional absorption spectroscopy.
% This anisotropic information is particularly useful for those samples that cannot be oriented conveniently in a polarized light beam (e.g., where single crystals can not be obtained). 
%
MCD spectra can be obtained from gases, solutions, or isotropic solids. Also,
MCD can be observed for any sample of molecules independent of whether they are chiral or not. One can use MCD to study molecules of high symmetry, and to probe degenerate electronic ground and excited states.

About fifty years ago, Buckingham and Stephens~\cite{ADB_and_Stephens}  described in an elegant and incisive way the theoretical foundation of MCD. For an electronic transition, the intensity of the signal is given by the contribution of three effects called ${\mathcal{A}}$, ${\mathcal{B}}$ and ${\mathcal{C}}$ terms. 
%According to eq.1 (Stephens' equation for MCD) 
The ${\mathcal{A}}$ term originates from the Zeeman splitting of degenerate excited states.
The ${\mathcal{B}}$ term arises from the mixing of the zero-field wavefunctions between nondegenerate states in the presence of a magnetic field. The ${\mathcal{C}}$ term is a temperature-dependent effect and originates from the Zeeman splitting of a degenerate ground state.
{Each term is associated with a characteristic band shape.}
After the seminal work of Buckingham and Stephens, the MCD spectra of several molecules were rationalized and understood qualitatively based on H{\"u}ckel molecular orbital, the Pariser- Parr-Pople (PPP) model\cch{,} and the  Complete Neglect of Differential Overlap/Spectroscopic
(CNDO/S) method.~\cite{Michl_semiempirico, castellan:6824, Meier}

The challenging aspect of the \textit{ab initio} computation of MCD spectra derives from the need to consider both the perturbation of a static magnetic field and the perturbation of an oscillating electric field.
In the last twenty years, several approaches have been proposed for the simulation of MCD, see e.g. Ref.~\citenum{MCD_Wires_review} for a review up to 2012.
% Briefly, some methods take into consideration all the three contributions (${\mathcal{A}}$, ${\mathcal{B}}$ and ${\mathcal{C}}$ terms), whereas others are focused only on one or two terms of MCD signal.
%
Among them, response theory~\cite{Response1,ChemRev_response,Norman:RespProp:2011} has been 
employed to formulate MCD in different forms, for instance as single residue of dipole-dipole-magnetic quadratic response functions,~\cite{MCD_Coriani_1999} as a complex polarization propagator,~\cite{Solheim2008, Seth2008:a} or a damped response function,~\cite{MCD_Thomas_damped} 
to avoid divergences,
and by magnetically-perturbed time-dependent density functional theory (MP-TDDFT) evaluating the perturbations induced into TDDFT
excitation energies and transition densities
%\revS{what?} 
by a static magnetic field.~\cite{Seth2008:a, Seth2008:b}
In the complex polarization propagator/damped response
framework, the MCD signal is computed directly, without
separation into MCD terms.~\cite{Solheim2008:a}
MCD spectra have also been calculated with sum-over-states (SOS) methods for the individual terms at the Hartree-Fock and DFT levels~\cite{Stepanek2013} of theory and within full configuration interaction (CI).~\cite{Honda2005} 
For the treatment of MCD arising from transition metals, DFT and HF may be  inadequate, thus multi-configurational self-consistent-field with the treatment of spin-orbit coupling (SOC) and spin-spin coupling (SSC) using complete active space self-consistent field (CASSCF)~\cite{Ganyushin_and_Neese} and restricted active space (RAS)~\cite{Autschbach_RAS} wavefunctions have been implemented.
% \revS{STRESS Neese's approach also gives THE CROSS SECTION WITHOUT SPLITTING INTO TERMS}
Gauge-origin independent formulations of MCD using the perturbative approach with London orbitals have been developed within DFT,~\cite{Krykunov2007, Kjaergaard2009} Hartree-Fock,~\cite{Kjaergaard2009} and coupled cluster (CC) frameworks.~\cite{MCD_CC, Kjaergaard2007}
Calculations of MCD within a variational 
treatment of the magnetic field have also been proposed.~\cite{Lee2011,Sun2019}

In this work we re-analyze the derivation of the MCD ${\mathcal{B}}$ 
term within resonant CC response function theory
and extend the theory
to the computation of the ${\mathcal{A}}$ term. 
{Then, we derive the} CC damped response expression for the MCD ellipticity. 
{Compared to the computation of induced transition strengths for stick spectra, the calculation of the damped response function is computationally more efficient for large chromophores or spectral regions with a high density of states.~\cite{Norman:RespProp:2011} In these cases the computation of the stick spectra requires the convergence of eigenvectors, and the calculation of (derivatives of)  transition moments for many states. The costs for the calculations of the damped response function depends mainly on the size of the frequency range and the frequency resolution, but is almost insensitive to the density of states. }
{To show the equivalence of the two approaches,} illustrative numerical results are reported at the coupled cluster singles and doubles (CCSD) level for the molecular systems cyclopropane and urea.
{These} are compared with TDDFT
(CAM-B3LYP) results.

\section{\label{sec:Theory}Theory}

%\subsection{The MCD cross section}
\subsection{\label{sec:standard}${\mathcal{A}}$ and ${\mathcal{B}}$  terms from resonant CC response theory}
% The anisotropy of the molar decadic extinction coefficients
% (in units of [M$^{-1}$cm$^{-1}$T$^{-1}$]) in an MCD experiment, where the strength of
% the external magnetic field is denoted by $B$, can be
% written as\cite{rizzo2012}
% %\revS{I AM MOVING THE SIGN INTO THETA!!}
% \begin{equation}
%     \frac{\Delta \epsilon(\omega)}{B_{ext}} = 
%     \frac{8 \pi^2 N_A}{
%         3 \times 1000 \times \ln(10) (4\pi\epsilon_0)\hbar c_0
%     } \ \theta_{{MCD}}
%     \label{MCD_Deltaepsi}
% \end{equation}
% where
Following Ref.~\citenum{MCD_Thomas_damped}, we write the 
ellipticity $\theta$ of plane-polarized light traveling in the $Z$ direction of
a space-fixed frame
through a sample of randomly
moving molecules in the presence of a magnetic field directed
along $Z$ 
as
\begin{equation}
    \theta = \frac{1}{6}\mu_0 c l N B_z \theta_{{MCD}}
\end{equation}
%
% For plane-polarized light traveling in the Z direction of
% a space-fixed frame, the  θ of a sample of randomly
% moving molecules in the presence of a magnetic field directed
% along the Z axis is given by
%
where, in atomic units,
%
%\begin{widetext}
%\begin{multline}
\begin{equation}
    \theta_{{MCD}} =
    {-}\omega
    \sum_{f}
    \bigg\{
        %\frac{1}{\hbar}
        \frac{\partial g(\omega,\omega_f)}{\partial \omega} 
        \mathcal{A} (0 \to {f})
 %\\
        + g(\omega,\omega_f)
        %\left[
            \mathcal{B} (0 \to {f}) 
            %+\frac{\mathcal{C} (0 \to {f})}{kT}
        %\right] 
    \bigg\}
\label{MCD_theta}
%\end{multline}
\end{equation}
%\end{widetext}
%%%
% \begin{widetext}
% \begin{equation}
%     \Delta \epsilon(\omega) = 
%     - \frac{8 \pi^2 N_A}{
%         3 \times 1000 \times \ln(10) (4\pi\epsilon_0)\hbar c_0
%     } B \omega
%     \sum_{f}\left\{
%         \frac{1}{\hbar} \frac{\partial g(\omega,\omega_f)}{\partial \omega} 
%         \mathcal{A} (0 \to {f}) + g(\omega,\omega_f)
%         \left[
%             \mathcal{B} (0 \to {f}) + \frac{\mathcal{C} (0 \to {f})}{kT}
%         \right] 
%     \right\}
% \label{MCD_spec_from_ABC}
% \end{equation}
% \end{widetext}
%%%
%
% In the equations above, $N_A$ is Avogadro's number, $c_0$ the velocity of light in vacuo, $\omega$ is the circular frequency, $k$ is Boltzmann's constant, $T$ is the temperature,
% $B_{ext}$ is the strength of the external magnetic field, and  
% %The lineshape function
% $g(\omega,\omega_j)$ is a lineshape function.
In the equations above, $N$ is the number density, $c$ is the velocity of light in vacuo, 
$\mu_0$ is the permeability in vacuo, $l$ is the
length of the sample,
$\omega$ is the circular frequency, 
%$k$ is Boltzmann's constant, $T$ is the temperature,
$B_{z}$ is the strength of the external magnetic field, and  
%The lineshape function
$g(\omega,\omega_j)$ is a lineshape function.
%\revS
{
We adopt the sign convention 
used by Michl.~\cite{Michl_MCD_1978} Thus, the contribution to $\theta_{MCD}$ of a transition $0\to f$ can consists of a positive (when $B < 0$) or negative (when $B > 0$) band of absorption-like shape centered at the position of the absorption band.
If the transition is degenerate, the absorption-like band is superimposed to a s-like (dispersive) shape, centered at the position of the absorption band, with a positive wing at lower energies and a negative one at higher energies (when $A < 0$) or with a negative wing at lower energies and a positive one at higher energies (when $A > 0$).
}
% \revS{SONIA: the above to me appears consistent with our definitions of the broadenings later on}
Note that, in the 
%From now on, we will focus our attention on the determination of the
expression of the 
%temperature-independent part of the
MCD ellipticity $\theta_{{MCD}}$ in Eq.~\eqref{MCD_theta},
we have omitted the 
temperature-dependent term, proportional to $\frac{\mathcal{C} (0 \to {f})}{kT}$, as it only contributes for systems with a 
degenerate ground state.
%and report everything in atomic units.
% $g(\omega,\omega_j) = \frac{\gamma^2}{\pi[(\omega-\omega_f)^2 + \gamma^2]}$ 
%\subsection{\label{sec:standard}${\mathcal{A}}$ and ${\mathcal{B}}$  terms from CC response theory}

The spectral representation of the ${\mathcal{A}}$ term for a 
non-degenerate ground state $0$ is~\cite{ADB_and_Stephens,BarronBook}
%\begin{multline}
\begin{equation}
    \label{SOS_MCD_ATERM}
    \mathcal{A} (0 \to {f})= 
    \tfrac{1}{2} \varepsilon_{\alpha \beta \gamma}
    \sum_{f' \in \mathfrak{D}_f} \operatorname{Im} \left[
        \langle 0 | \mu_\alpha | f \rangle
        \langle f | m_\gamma | f' \rangle
        \langle f' | \mu_\beta | 0 \rangle
    \right]
    %\end{multline}
\end{equation}
where $\mu_\alpha$ and $\mu_\beta$ are components of the electric dipole operator,
$m_\gamma$ is a component of the magnetic dipole operator, 
and $\varepsilon_{\alpha \beta \gamma}$ is the Levi-Civita tensor. Implicit summation over repeated Greek indices is assumed.
$\mathfrak{D}_f$ is the set of degenerate states of which $f$ is a part.~\cite{ADB_and_Stephens} 
The $\mathcal{A}$ term vanishes for a non-degenerate excited state (as the magnetic moment is quenched).~\cite{Schatz_Book}
% whereas for degenerate excited states the contribution from  usually dominates.~\cite{REF}  
% \revS{SIMONE, CAN YOU CONFIRM THE STATEMENT ABOVE AND ADD A REFERENCE? Simone: Okay. reference added.
% WE NEED TO REPHRASE. IT IS NOT A as stick that is larger, }
% %$j'$ is summed over $j$ and all states degenerate with $j$.

The spectral representation of the ${\mathcal{B}}$ term is given 
by~\cite{ADB_and_Stephens,BarronBook}
\begin{multline}
    \label{SOS_MCD_STRESP}
    {\mathcal{B}} (0 \rightarrow{f})= \varepsilon_{\alpha\beta\gamma}
    \operatorname{Im} \Bigg[
        \sum_{k\neq 0}   \frac{\langle k|m_\gamma| 0 \rangle}{\omega_{k}}
        { \langle 0| \mu_\alpha | f \rangle}  
        {\langle f| \mu_\beta |k \rangle}
        + \sum_{k\notin \mathfrak{D}_f} 
        \frac{ \langle f| m_\gamma | k \rangle}{\omega_{k} - \omega_{f}}
        { \langle 0| \mu _\alpha |f \rangle}
        {\langle k | \mu _\beta | 0\rangle}
    \Bigg]
\end{multline}
A connection has previously been made between the $\mathcal{B}$ term of a non degenerate state and the 
derivative of the transition strength matrix.~\cite{MCD_CC}
%\begin{equation}
%    {\mathcal{B}}(0\to f) = \tfrac{1}{2} \varepsilon_{\alpha\beta\gamma}
%    \operatorname{Im}
%    \left(
%        \frac{dS_{of}^{\mu_\alpha\mu_\beta}}{dB_\gamma}
%    \right)
%\end{equation}
We will here extend this definition in order to include the $\mathcal{A}$-term.
For exact states, the magnetic-field derivative of the electric-dipole transition strength, $S_{of}^{\mu_\alpha\mu_\beta}={ \langle 0| \mu_\alpha | f \rangle}  
        {\langle f| \mu_\beta |0 \rangle}$, is
\begin{multline}
    \tfrac{1}{2}
    \operatorname{Im}
    \left(
        \frac{dS_{of}^{\mu_\alpha\mu_\beta}}{dB_\gamma}
    \right) =
    \operatorname{Im} \Bigg[
        \sum_{k\neq 0}   \frac{\langle k|m_\gamma| 0 \rangle}{\omega_{k}}
        { \langle 0| \mu_\alpha | f \rangle}  
        {\langle f| \mu_\beta |k \rangle} 
        %\\
        + \sum_{k \neq f} 
        \frac{ \langle f| m_\gamma | k \rangle}{\omega_{k} - \omega_{f}}
        { \langle 0| \mu _\alpha |f \rangle}
        {\langle k | \mu _\beta | 0\rangle}
    \Bigg]
\end{multline}
which is exactly the expression for the contributions to the $\mathcal{B}$ term
if the state $f$ is non-degenerate.
If $f$ is degenerate, however, the second sum contains additional terms, explicitly excluded from Eq.~\eqref{SOS_MCD_STRESP},
involving the states degenerate with the final state.
If we assume that the degeneracy can be broken by an infinitesimal amount,
$\eta = \omega _{f'} - \omega_f$, the $\mathcal{A}$ term can be defined
as the residue
\begin{equation}
\label{A_residue_exact}
    \mathcal{A} (0 \rightarrow{f})=
    \tfrac{1}{4} \varepsilon_{\alpha \beta \gamma}
    \lim _{\eta \to 0} \eta 
    \operatorname{Im}
    \left(
        \frac{dS_{0f}^{\mu_\alpha\mu_\beta}}{dB_\gamma}
    \right)
\end{equation}
Similarly, the expression for the $\mathcal{B}$ term in 
Eq.~\eqref{SOS_MCD_STRESP} is
obtained %if we define 
by defining
the $\mathcal{B}$ term as what remains of the transition-moment derivative once the singularities are removed, i.e.
any degeneracy is projected out of the excited state wavefunction response.

In CC response theory, the transition strength is given as the product of
distinct left and right transition moments~\cite{christiansen1998,cc-multiphoton-transition-moment,MCD_Coriani_1999,ChemRev_response}
\begin{align}
    S_{0f}^{\mu_\alpha\mu_\beta} &= 
        \tfrac{1}{2} T_{0f}^{\mu_\alpha} T_{f0}^{\mu_\beta} +
        \tfrac{1}{2}(T_{0f}^{\mu_\beta} T_{f0}^{\mu_\alpha})^*, \\
    T_{0f}^{\mu_\alpha} &= \lhs{\mu_\alpha} \rightv{f} +
                           {{M}}_f \rhs{\mu_\alpha}, \\
    T_{f0}^{\mu_\beta} &= \leftv{f} \rhs{\mu_\beta}.
\end{align}
The eigenvectors are obtained {by} solving the right and left eigenvalue equations
\begin{align}
    (\textbf{A}-\omega_{f}\textbf{1})\rightv{f} = \textbf{0}\\
    \leftv{f}(\textbf{A}-\omega_{f}\textbf{1}) = \textbf{0}
\end{align}
under the biorthogonality condition ${L}_{k}{R}_{l}=\delta_{lk}$, and the transition multipliers $M_f(\omega_f)$ are
the solution of the linear equation
\begin{equation}
    M_f(\omega_f) (\textbf{A}+\omega_{f}\textbf{1})=- \textbf{F}\rightv{f}~~.
\end{equation}
For ease of notation we have omitted the overbar {on} the transition multiplier. The definition{s} of the Jacobian matrix $\textbf{A}$, the matrix $\textbf{F}$
and of the property gradients, $\rhs{X}$ and $\lhs{X}$,
for any generic operator $X$,
can be found, e.g., in Ref.~\citenum{christiansen1998}.

%In response theory terms it the $\mathcal{A}$ and $\mathcal{B}$ terms
%arise for double and single residues of the quadratic response function, and that is how we will distinguish them in CC theory.
%{\bf{Add expressions for S in terms of left and right TM.}}

Let us start by considering the case where the final state $f$ is not degenerate.
Straightforward differentiation of the CC left and right ground-to-excited-state transition moments yields
\begin{align}
\label{dleftmom}
    \frac{dT_{0f}^{\mu_\alpha}}{dB_\gamma} &=
    -T_{0f}^{\mu_\alpha m_\gamma} 
    = \lhs{\mu_\alpha} \rightv{f}^{m_\gamma} + 
(\mathbf{F}^{\mu_\alpha} t^{m_\gamma} + \bar{t}^{m_\gamma} \mathbf{A}^{\mu_\alpha} ) \rightv{f}
        + M_f^{m_\gamma} \rhs{\mu_\alpha} + M_f \mathbf{A}^{\mu_\alpha} t^{m_\gamma} \\
    \frac{dT_{f0}^{\mu_\beta}}{dB_\gamma} &=
    -T_{f0}^{\mu_\beta m_\gamma} 
    = \leftv{f}^{m_\gamma} \rhs{\mu_\beta} + \leftv{f} \mathbf{A}^{\mu_\beta} t^{m_\gamma}
\label{drightmom}
\end{align}
where $t^{m_\gamma}$ and $\bar{t}^{m_\gamma}$ are the zero-frequency
derivatives, with respect to the magnetic field, of the CC amplitudes and Lagrangian multipliers, respectively, obtained solving usual right and left response equations:
\begin{align}
  (\mathbf{A}-\omega_A {\bf{1}}) t^{X} (\omega_X) &= -\xi^{X}\\\nonumber
  \bar{t}^{X}(\omega_X) (\mathbf{A}+\omega_X {\bf{1}})  
  &= -(\eta^{X}+{\bf{F}}t^{X} (\omega_X))~\\
  &= -\bar{\xi}^X(\omega_X)
\end{align}
for operator $X$ equal to $m_\gamma$ and $\omega_X = 0$.

The equations determining the magnetic-field
derivatives,
$\leftv{f}^{m_\gamma}$ and
$\rightv{f}^{m_\gamma}$, 
of the left and right eigenvectors,
as well as the magnetic-field derivative ${{M}}_f^{m_\gamma}$ of 
% the response
% \revS{excited state} 
the transition
multipliers, are
\begin{align}
    \label{deriv_r_vec}
    (\mathbf{A} -\omega_f\mathbf{1}) \rightv{f}^{m_\gamma} =& 
    - \left( 
        \mathbf{A}^{m_\gamma}
        +\mathbf{B} t^{m_\gamma} - \omega_f^{m_\gamma} \mathbf{1}
    \right) \rightv{f} , \\
    \label{deriv_l_vec}
    \leftv{f}^{m_\gamma} (\mathbf{A} -\omega_f \mathbf{1}) =& 
    - \leftv{f} \left( 
        \mathbf{A}^{m_\gamma}
        + \mathbf{B} t^{m_\gamma} - \omega_f^{m_\gamma}\mathbf{1}
    \right), \\
    M_f^{m_\gamma} ( \mathbf{A} + \omega _{f}\mathbf{1} ) =& 
            - \mathbf{F} \rightv{f}^{m_\gamma} 
            - (\mathbf{F}^{m_\gamma} + \mathbf{G} t^{m_\gamma} + \bar{t} ^{m_\gamma} \mathbf{B}  ) \rightv{f} \label{deriv_m_vec}
            - M_f ( \mathbf{A}^{m_\gamma} + \omega_f^{m_\gamma}\mathbf{1} + \mathbf{B} t^{m_\gamma} )
\end{align}
where   
\begin{equation}\omega_f^{m_\gamma} = \leftv{f} ( \mathbf{A}^{m_\gamma} + \mathbf{B} t^{m_\gamma}) \rightv{f}~.
\end{equation}
See again Ref.~\citenum{christiansen1998} for the definition of the remaining CC matrices.

While $(\mathbf{A} -\omega_f\mathbf{1})$ in 
Eqs.~\eqref{deriv_r_vec} and~\eqref{deriv_l_vec}
is singular, it is easy to show that 
the right hand sides are orthogonal to $\rightv{f}$ and $\leftv{f}$, respectively.
It is sufficient to insert $\omega_f^{m_\gamma}$ in their definition and to project them against $\leftv{f}$ and $\rightv{f}$, respectively.
Thus, for non-degenerate final states $f$,  
Eqs.~\eqref{deriv_r_vec} and \eqref{deriv_l_vec}
can be solved in the orthogonal complement to the singularity without loss of generality.~\cite{MCD_CC,cc-multiphoton-transition-moment,MPTM_CC} 
In practice,
this is achieved by 
introducing the projector
\begin{equation}
      P_f = 1 - \rightv{f} \leftv{f} 
\end{equation}
% To this end, we introduce the projector
% \begin{equation}
% \label{degP}
%      P_f = 1 - \sum_{f' \in \mathfrak{D}_f} \rightv{f'} \leftv{f'} 
% \end{equation}
and the projected derivative eigenvectors
\begin{align}
    {}^\perp \rightv{f}^{m_\gamma} =& P_f \rightv{f}^{m_\gamma} \\
    {}^\perp \leftv{f}^{m_\gamma} =& \leftv{f}^{m_\gamma} P_f
\end{align}
which are obtained solving
\begin{align}
\label{Perp_Rfm}
    P_f (\mathbf{A} -\omega_f) {}^\perp \rightv{f}^{m_\gamma} =& 
    - P_f \left(
        \mathbf{A}^{m_\gamma}
        +\mathbf{B} t^{m_\gamma}
    \right) \rightv{f} , \\
    \label{Perp_Lfm}
    {}^\perp \leftv{f}^{m_\gamma} (\mathbf{A} -\omega_f) P_f =& 
    - \leftv{f} \left( 
        \mathbf{A}^{m_\gamma}
        + \mathbf{B} t^{m_\gamma}
    \right) P_f
    .
    % , \\
    % {}^\perp M_f^{m_\gamma} ( \mathbf{A} + \omega _{f} ) =& 
    %         - \mathbf{F} {}^\perp \rightv{f}^{m_\gamma} 
    %         - (\mathbf{F}^{m_\gamma} + \mathbf{G} t^{m_\gamma} + \bar{t} ^{m_\gamma} \mathbf{B}  ) \rightv{f} 
    %         \nonumber \\&  
    %         - M_f ( \mathbf{A}^{m_\gamma} + \omega_f^{m_\gamma} + \mathbf{B} t^{m_\gamma} ).
\end{align}
In addition, we use the notation 
${}^\perp M_f^{m_\gamma}$ to emphasize that 
the Lagrange multiplier responses
are calculated using the non-singular derivative of the eigenvector, i.e.
\begin{multline}
\label{perp_Mfm}
    {}^\perp M_f^{m_\gamma} ( \mathbf{A} + \omega _{f} ) =
            - \mathbf{F} {}^\perp \rightv{f}^{m_\gamma} 
            - (\mathbf{F}^{m_\gamma} + \mathbf{G} t^{m_\gamma} + \bar{t} ^{m_\gamma} \mathbf{B}  ) \rightv{f}
            - M_f ( \mathbf{A}^{m_\gamma} + \omega_f^{m_\gamma} + \mathbf{B} t^{m_\gamma} )
\end{multline} 

If the final state $f$ is degenerate (i.e., it belongs to the set $\mathfrak{D}_f$), 
the projector is generalized as
\begin{equation}
\label{degP}
     P_f = 1 - \sum_{f' \in \mathfrak{D}_f} \rightv{f'} \leftv{f'}~. 
\end{equation}
%and the projected derivative eigenvectors
%\begin{align}
%    {}^\perp \rightv{f}^{m_\gamma} =& P_f \rightv{f}^{m_\gamma} \\
%    {}^\perp \leftv{f}^{m_\gamma} =& \leftv{f}^{m_\gamma} P_f
%\end{align}
%which are obtained solving
%\begin{align}
%\label{Perp_Rfm}
%    P_f (\mathbf{A} -\omega_f) {}^\perp \rightv{f}^{m_\gamma} =& 
%    - P_f \left(
%        \mathbf{A}^{m_\gamma}
 %       +\mathbf{B} t^{m_\gamma}
 %   \right) \rightv{f} , \\
  %  \label{Perp_Lfm}
   % {}^\perp \leftv{f}^{m_\gamma} (\mathbf{A} -\omega_f) P_f =& 
%    - \leftv{f} \left( 
 %       \mathbf{A}^{m_\gamma}
  %      + \mathbf{B} t^{m_\gamma}
   % \right) P_f
    %.
    % , \\
    % {}^\perp M_f^{m_\gamma} ( \mathbf{A} + \omega _{f} ) =& 
    %         - \mathbf{F} {}^\perp \rightv{f}^{m_\gamma} 
    %         - (\mathbf{F}^{m_\gamma} + \mathbf{G} t^{m_\gamma} + \bar{t} ^{m_\gamma} \mathbf{B}  ) \rightv{f} 
    %         \nonumber \\&  
    %         - M_f ( \mathbf{A}^{m_\gamma} + \omega_f^{m_\gamma} + \mathbf{B} t^{m_\gamma} ).
%\end{align}
Then, we introduce a distinction between the two kinds of
contributions, i.e., the $\mathcal{A}$ and the $\mathcal{B}$ term:
In accordance with exact theory, we define the $\mathcal{B}$ term as the term obtained by projecting out the singularity and otherwise continuing as in the non-degenerate case. 
The $\mathcal{A}$ term, on the other hand, will be defined as the residue of the term involving the singularity.
Thus, the CC $\mathcal{B}$ term will be obtained 
as
\begin{equation}
    \label{MCD_Bterm_CC}
    \mathcal{B}_\mathrm{CC}(0\to f) = 
    - \tfrac{1}{2} \varepsilon_{\alpha\beta\gamma} 
    \left(
        {}^\perp T^{\mu_\alpha m_\gamma}_{0f} T^{\mu_\beta}_{f0} +
        T^{\mu_\alpha}_{0f} {}^\perp T^{\mu_\beta m_\gamma}_{f0}
    \right)
\end{equation}
where the perpendicular perturbed 
transition moments (${}^\perp T^{\mu_\alpha m_\gamma}_{0f}$ and
${}^\perp T^{\mu_\beta m_\gamma}_{f0}$) are defined by 
introducing ${}^\perp \rightv{f}^{m_\gamma}$, ${}^\perp \leftv{f}^{m_\gamma}$ 
and ${}^\perp M_f ^{m_\gamma}$
in place of their non-$\perp$ 
equivalents into
Eqs.~\eqref{dleftmom} and~\eqref{drightmom}. 
%leads to an expression for
%the $\mathcal{B}$ term which is well-defined for degenerate states and
%reduce to the well known description for non-degenerate states.
The formulation of the derivative transition moments as in Eqs. \eqref{dleftmom} and \eqref{drightmom} is attractive as 
all dependencies on the electric dipole components
$\mu_\alpha$ and $\mu_\beta$ are explicit, allowing for the identification of derivative left and right transition densities.

An alternative expression of the ({orthogonal})
left moment 
is obtained by eliminating $M_f^{m_\gamma}$ 
(or ${}^\perp M_f ^{m_\gamma}$) from Eq.~\eqref{dleftmom} using Eq.~\eqref{deriv_m_vec} (or Eq~\eqref{perp_Mfm})
\begin{align}
    M_f^{m_\gamma} \rhs{\mu_\alpha} & = 
    - \Big[
            \mathbf{F} \rightv{f}^{m_\gamma} 
            + (\mathbf{F}^{m_\gamma} + \mathbf{G} t^{m_\gamma} + \bar{t} ^{m_\gamma} \mathbf{B}  ) \rightv{f} 
            \nonumber \\&  
            \quad + M_f ( \mathbf{A}^{m_\gamma} + \omega_f^{m_\gamma}\mathbf{1} + \mathbf{B} t^{m_\gamma} )
            \Big]
 \Big(\mathbf{A}+\omega_f\mathbf{1}\Big)^{-1} \xi^{\mu_\alpha}
    % -\Big[\mathbf{F}R^{m_\gamma}_f \\
    %  + \big(\mathbf{F}^{m_\gamma} + \mathbf{G} t^m + \bar{t}^m\mathbf{B}\big) R_f + M_f\big(\mathbf{A}^m+\omega^m_f \mathbf{1} + \mathbf{B} t^m\big) \Big] \Big[\mathbf{A}+\omega_f\mathbf{1}\Big]^{-1} \xi^d
    \nonumber \\ 
    & = \Big[
            \mathbf{F} \rightv{f}^{m_\gamma} 
            + (\mathbf{F}^{m_\gamma} + \mathbf{G} t^{m_\gamma} + \bar{t} ^{m_\gamma} \mathbf{B}  ) \rightv{f} 
            \nonumber \\&  
            \quad + M_f ( \mathbf{A}^{m_\gamma} + \omega_f^{m_\gamma} \mathbf{1}+ \mathbf{B} t^{m_\gamma} )
            \Big] \ampd{\mu_\alpha}{-\omega_f},
\end{align}
% yielding \revS{(I NEED TO RECHECK SIGN AND $\perp$ SYMBOLS)}
%This is now rearranged to:
\begin{align}
\nonumber
    T^{\mu_\alpha m_\gamma}_{0f} = & \big[\mathbf{G}t^{m_\gamma} t^{\mu_\alpha}(-\omega_f) + \mathbf{F}^{m_\gamma} t^{\mu_\alpha}(-\omega_f) + \mathbf{F}^{\mu_\alpha} t^{m_\gamma} \big] R_f
    \\  &  \nonumber
    + M_f \big[\mathbf{A}^{\mu_\alpha} t^{m_\gamma} + \mathbf{A}^{m_\gamma} t^{\mu_\alpha}(-\omega_f) + \mathbf{B} t^{m_\gamma} t^{\mu_\alpha}(-\omega_f) \big]
 \\ & \nonumber
   + \big[\eta^{\mu_\alpha} + \mathbf{F} t^{\mu_\alpha}(-\omega_f)\big] R^{m_\gamma}_f 
  \\ & \nonumber
   + \omega^{m_\gamma}_f \cdot M_f t^{\mu_\alpha}(-\omega_f) 
   \\ & 
 + \bar{t}^{m_\gamma} \big[\mathbf{A}^{\mu_\alpha} + \mathbf{B} t^{\mu_\alpha}(-\omega_f)\big] R_f 
 \label{left_v1}
%  \nonumber
%  \\ 
%   = &  \big[\mathbf{G}t^{m_\gamma} t^{\mu_\alpha}(-\omega_f) + \mathbf{F}^m t^d(-\omega_f) + \mathbf{F}^{\mu_\alpha} t^{m_\gamma}\big] R_f
%     \\  &  \nonumber
%     + M_f \big[ \mathbf{A}^{\mu_\alpha} t^{m_\gamma} + \mathbf{A}^{m_\gamma} t^{\mu_\alpha}(-\omega_f) + \mathbf{B} t^{m_\gamma} t^{\mu_\alpha}(-\omega_f) \big]
%  \\ & \nonumber
%   + \big(\eta^{\mu_\alpha} + \mathbf{F} t^{\mu_\alpha}(-\omega_f)\big) R^{m_\gamma}_f 
%   \\ & \nonumber
%   + \omega^{m_\gamma}_f \cdot  M_f t^{\mu_\alpha}(-\omega_f)
%   \\ & \nonumber
%  + \big(\eta^{m_\gamma} + \mathbf{F} t^{m_\gamma}\big) R^{\mu_\alpha}_f(-\omega_f) 
\end{align}
The last term in Eq.~\eqref{left_v1} can be further
replaced by 
%\mbox{
\begin{equation}
\big(\eta^{m_\gamma} + \mathbf{F} t^{m_\gamma}\big) R^{\mu_\alpha}_f(-\omega_f)
= {\bar{\xi}}^{m_\gamma} (0) R^{\mu_\alpha}_f(-\omega_f)
\end{equation}
%},
which now involves  $R^{\mu_\alpha}_f(-\omega_f)$, the first-order response to the electric field of the right eigenvector in a non-phase-isolated (i.e. unprojected) form.~\cite{MPTM_CC}
Similarly, the third term can be recast as
\begin{equation}
\label{eq31}
\big[\eta^{\mu_\alpha} + \mathbf{F} t^{\mu_\alpha}(-\omega_f)\big] {}^\perp R^{m_\gamma}_f
=  -{}^\perp \bar{t}(-\omega_f) \left( 
        \mathbf{A}^{m_\gamma}
        +\mathbf{B} t^{m_\gamma} - \omega_f^{m_\gamma} \mathbf{1}
    \right) \rightv{f}
\end{equation}
% , i.e.\ where the contribution parallel to the unperturbed eigenvectors are not projected out.
%
Eq.~\eqref{left_v1} is {formally} the approach taken in the implementation in Dalton~\cite{MCD_CC,DaltonPaper}
and Turbomole,~\cite{MCD_Khani,turbomole2}
the latter also employing Eq.~\eqref{eq31}.~\cite{MCD_Khani,turbomole2}
% \revS{SONIA: does turbomole solve for $R^{\mu_\alpha}_f(-\omega_f)$?}
% \textcolor{red}{Christof: yes, it does.}
If the final states $f$ are non-degenerate, both approaches (Eq.~\eqref{dleftmom} and  \eqref{left_v1})
require the solution of the same amount of linear equations.
In the case of degenerate states, however, the latter is advantageous 
as the dipole response amplitudes $\ampd{\mu_\alpha}{-\omega_f}$ need
to be calculated only once for each 
{degenerate} set.
%TO RASMUS!!!!
% \textcolor{red}{TO RASMUS: please explain more clearly which costs are saved.}

To obtain the CC expression for the $\mathcal{A}$ term, 
we %once again 
perform a residue analysis according to Eq.~\eqref{A_residue_exact}, i.e. 
\begin{equation}
    \mathcal{A}_\mathrm{CC} (0 \rightarrow{f})= 
    -\tfrac{1}{4} \varepsilon_{\alpha \beta \gamma}
    \lim _{\eta \to 0} \eta 
    \operatorname{Im}
    \left(
        T^{\mu_\alpha m_\gamma}_{0f} T^{\mu_\beta}_{f0} +
        T^{\mu_\alpha}_{0f} T^{\mu_\beta m_\gamma}_{f0}
    \right)
\end{equation}
which requires the residues
\begin{align}
    {}^\parallel\!T_{0f}^{\mu_\alpha m_\gamma} &= 
     \lim_{\eta \to 0} \ \eta \ T_{0f}^{\mu_\alpha m_\gamma} =
     - \lhs{\mu_\alpha} \ {}^\parallel\!\rightv{f}^{m_\gamma} 
     - {}^\parallel\!M_f^{m_\gamma} \ \rhs{\mu_\alpha}~, \\
    {}^\parallel\!T_{f0}^{\mu_\beta m_\gamma} &= 
     \lim_{\eta \to 0} \ \eta \ T_{f0}^{\mu_\beta m_\gamma} =
     - {}^\parallel\!\leftv{f}^{m_\gamma} \rhs{\mu_\beta}~.
\end{align}
The response of the eigenvectors parallel to the degenerate set $\mathcal{D}_f$ are defined as residues of non-phase isolated derivatives 
of the eigenvectors~\cite{MPTM_CC}
\begin{equation}
    {}^\parallel\!\rightv{f}^{m_\gamma} =
    \lim_{\eta \to 0} \ \eta \ \rightv{f}^{m_\gamma} =
    %\\
    -\hspace{-12pt}
    %\sum_{f'\in \mathfrak{D}_f, f' \neq f} 
    \sum_{
    \scriptsize{
    \begin{array}{c}
    f'\in \mathfrak{D}_f, \\
    f' \neq f
    \end{array} }}
    \hspace{-12pt}
    \rightv{f'} \leftv{f'} ( 
        \mathbf{A}^{m_\gamma} + \mathbf{B} t^{m_\gamma}
    ) \rightv{f}
= - \hspace{-12pt}
\sum_{
    \scriptsize{
    \begin{array}{c}
    f'\in \mathfrak{D}_f, \\
    f' \neq f
    \end{array} }}
     \hspace{-6pt}
\rightv{f'} T^{m_\gamma}_{f'f}
\end{equation}
\begin{equation}
    {}^\parallel\!\leftv{f}^{m_\gamma} =
    \lim_{\eta \to 0} \ \eta \ \leftv{f}^{m_\gamma} = %\\
    -\hspace{-12pt}
    %\sum_{f'\in \mathfrak{D}_f, f' \neq f} 
    \sum_{
    \scriptsize{
    \begin{array}{c}
    f'\in \mathfrak{D}_f, \\
    f' \neq f
    \end{array} }}
    \hspace{-12pt}
    \leftv{f} ( 
        \mathbf{A}^{m_\gamma} + \mathbf{B} t^{m_\gamma}
    ) \rightv{f'} \leftv{f'}
    = 
    -\hspace{-12pt}
    %\sum_{f'\in \mathfrak{D}_f, f' \neq f} 
    \sum_{
    \scriptsize{
    \begin{array}{c}
    f'\in \mathfrak{D}_f, \\
    f' \neq f
    \end{array} }}
    \hspace{-6pt}
    T^{m_\gamma}_{ff'} \leftv{f'}
    \end{equation}
    and
\begin{align}
    {}^\parallel\!M_f^{m_\gamma} &=
    - \mathbf{F}\ {}^\parallel\!\rightv{f}^{m_\gamma}
    %(\mathbf{A} + \gamma_f)^{-1} ????
    (\mathbf{A} + \omega_f {\bf{1}})^{-1}
    = 
    -\hspace{-12pt}
    \sum_{
    \scriptsize{
    \begin{array}{c}
    f'\in \mathfrak{D}_f, \\
    f' \neq f
    \end{array} }}
    \hspace{-5pt}
    M_{f'} T^{m_\gamma}_{f'f}~,
\end{align}
In the equations above, 
simplifications have been made by identifying the conventional
CC expression for transition moments between excited states, e.g.
$T_{ff'}^{m_\gamma} = \leftv{f} ( 
        \mathbf{A}^{m_\gamma} + \mathbf{B} t^{m_\gamma}
    ) \rightv{f'}$.
This allows us to write the $\mathcal{A}_\mathrm{CC}$ term as
\begin{equation}
    \mathcal{A}_\mathrm{CC} (0 \rightarrow{f})= 
    -\tfrac{1}{4} \varepsilon_{\alpha \beta \gamma}
    \operatorname{Im}
    \hspace{-8pt}
        \sum_{
    \scriptsize{
    \begin{array}{c}
    f'\in \mathfrak{D}_f, \\
    f' \neq f
    \end{array} }}
    \hspace{-8pt}
    %\sum_{f'\in \mathfrak{D}_f, f' \neq f} 
    \left(
        T^{\mu_\alpha}_{0f'} T^{m_\gamma}_{f'f} T^{\mu_\beta}_{f0} +
        T^{\mu_\alpha}_{0f} T^{m_\gamma}_{ff'} T^{\mu_\beta}_{f'0}
    \right)
\end{equation}
or, when summed over the whole degenerate set, 
%(which symbol to use?)
\begin{equation}
    \label{MCD_Aterm_CC}
    \mathcal{A}_\mathrm{CC} (0 \rightarrow{\mathfrak{D}_f})= 
    - \tfrac{1}{2} \varepsilon_{\alpha \beta \gamma}
    \operatorname{Im}
    \hspace{-8pt}
    \sum_{f', f'' \in \mathfrak{D}_f} 
    \hspace{-8pt}
    (1 - \delta_{f'f''})
    \left(
        T^{\mu_\alpha}_{0f'} T^{m_\gamma}_{f'f''} T^{\mu_\beta}_{f''0}
    \right)~.
\end{equation}

{Note that the $\mathcal{A}$ term has previously been formulated as the derivative of the excitation
frequency~\cite{Krykunov2007,MCD_Thomas_damped,BarronBook}
%[in accordance with time-dependent perturbation
%theory
\begin{equation}
    \mathcal{A} (0 \to {f})=
    -\tfrac{1}{2} \varepsilon_{\alpha \beta \gamma}
    \sum_{f\in \mathfrak{D}_f} \left(\frac{\partial \omega_{f}}{\partial B_\gamma}\right)
    \operatorname{Im}\left\{
    \mu_\alpha^{0\tilde{f}}\mu_\beta^{\tilde{f}0}
    \right\}
\end{equation}
where 
the real degenerate
states $f$ are (typically) expanded in complex states $\tilde{f}$, which diagonalize the imaginary
operator $m_\gamma$.~\cite{Krykunov2007,MCD_Thomas_damped}
This is consistent with our derivation, as we can identify
\begin{equation}
\frac{\partial \omega_{{f}}}{\partial B_\gamma}
= L_f(\mathbf{A}^{m_\gamma} + \mathbf{B} t^{m_\gamma})R_{f'} 
%\equiv 
= T_{f {f'}}
\end{equation}
% DOES THIS MAKES THE ENTIRE DERIVATION EARLIER A BIT REDUNDANT?
% CAN WE SAY THIS:
Our derivation highlights how the transformation to the diagonal basis for $m_\gamma$ can be avoided.
}

% \revS{SONIA TO CH + RF: Can one introduce $N^{ff'}(0)$ vectors to lower the order of perturbed eqs to solve?
% RF: $t^{m_\gamma}$ is already needed in the B-term, and even if it was not, it is only three vectors, whereas $N^{ff'}$ would be two per set of doubly degenerate states, so I don't think that would save anything.
% }

\subsection{MCD spectra from CC damped response theory}
\label{sec:damped}

% \revS{SONIA 2 RASMUS: I am changing the name of the broadening parameter to $\varpi$ to avoid confusion with the $\gamma$ direction.}

Within damped response theory, the MCD ellipticity can be obtained directly from the magnetic field derivative of the damped polarizability:
%\revS{What is the meaning of the pedix $\gamma$ on $\theta$?}
%REF!!:
\begin{equation}
    %\theta _{MCD, \gamma} = - \; \omega %\! \! \! \!
    \theta _{{MCD}} = - \; \omega %\! \! \! \! %\sum_{\alpha,\beta,\gamma = x, y, z}
    %\! \! \!
        \epsilon _{\alpha\beta\gamma} %B_\gamma
        \operatorname{Re} \left(
        \frac{
            d \langle\langle \mu_\alpha; \mu_\beta \rangle\rangle_{\omega+i\varpi}
        }{
            d B_\gamma
        }
        \right)_{B = 0} .
    \label{MCD_damped_def}
\end{equation}
In coupled cluster theory, the damped polarizability can be written as
given in Refs.~\citenum{KauczorCCCPP,FaberJCTC,FaberPCCP}:
\begin{align}
    \label{cmplxLR}
    \langle\langle \mu_\alpha; 
        \mu_\beta \rangle\rangle_{\omega+i\varpi}
    = \frac{1}{2}C^{\pm \omega}
    \big\{
        &\lhs{\mu_\alpha} \ampd{\mu_\beta}{\omega+i\varpi} +
        \\ \nonumber
        &\lhs{\mu_\beta} \ampd{\mu_\alpha}{-\omega-i\varpi}+
        \\ \nonumber
        &\mathbf{F}\ampd{\mu_\beta}{\omega+i\varpi}
            \ampd{\mu_\alpha}{-\omega-i\varpi}
    \big\} .
\end{align}
The complex amplitudes are found solving {the} complex linear equations:
\begin{equation}
\label{cmplx_eq}
 [\mathbf{A} -(\omega+i\varpi)\boldsymbol{1}]  \ampd{\mu_\alpha}{\omega+i\varpi} = -\rhs{\mu_\alpha} .
\end{equation}
We refer to our previous work~\cite{KauczorCCCPP,FaberJCTC,FaberPCCP} for details on how to solve the complex equations in Eq.~\eqref{cmplx_eq}. 
%   
% \revS{ALSO: please check what you want to do with the magnetic field strength $B_\gamma$... See also how I defined $\theta_{MCD}$ at the beginning. If we find a coherent definition, there is no need to redo the figures}

%\revR{Stuff about CC damped response in general.....}

Typically, the CC response functions need to be explicitly symmetrized,~\cite{christiansen1998}
as indicated in Eq.~\eqref{cmplxLR} by the $\frac{1}{2} C^{\pm\omega}$ operator.
However, the Levi-Civita symbol in Eq.~\eqref{MCD_damped_def} makes this symmetrization redundant.
Taking the first derivative of {the} non-symmetric CC linear response function, i.e. the term in brackets in Eq.~\eqref{cmplxLR}, we obtain:
%\revS{Explain why we choose the asymmetric}
\begin{multline}
     \frac{
         d \langle\langle \mu_\alpha; \mu_\beta \rangle\rangle_{\omega+i\varpi}
         }{
         d B_\gamma} =  \;
     \mathbf{F}^{m_\gamma} \ampd{\mu_\beta}{\omega+i\varpi} \ampd{\mu_\alpha}{-\omega-i\varpi}
     \\+
     \Big[
 \mathbf{F}^{\mu_\alpha} \ampd{\mu_\beta}{\omega+i\varpi} +
\mathbf{F}^{\mu_\beta}\ampd{\mu_\alpha}{-\omega-i\varpi} 
     \\
     + \mathbf{G} \ampd{\mu_\beta}{\omega+i\varpi} \ampd{\mu_\alpha}{-\omega-i\varpi}
     \Big] t^{m_\gamma} 
     \\+ 
 \bar{t}^{m_\gamma} \Big[
       \mathbf{A}^{\mu_\alpha} \ampd{\mu_\beta}{\omega+i\varpi} +
       \mathbf{A}^{\mu_\beta}
       \ampd{\mu_\alpha}{-\omega-i\varpi}
     \\+ 
         \mathbf{B} \ampd{\mu_\beta}{\omega+i\varpi} \ampd{\mu_\alpha}{-\omega-i\varpi}
     \Big] 
     \\+ 
     \left[
         \mathbf{F} \ampd{\mu_\alpha}{-\omega-i\varpi} + \lhs{\mu_\alpha}
    \right] \ampd{\mu_\beta m_\gamma}{\omega+i\varpi}
     \\ +
     \left[
         \mathbf{F} \ampd{\mu_\beta}{\omega+i\varpi} + \lhs{\mu_\beta} 
     \right] \ampd{\mu_\alpha m_\gamma}{-\omega-i\varpi}
     \label{eq.dCPPLR}
\end{multline}
The above expression contains the 
%second derivatives of the amplitudes, 
doubly perturbed amplitudes,
which are defined by the second-order response equations 
\begin{multline}
        \left[
        \mathbf{A} + (\omega+i\varpi)
    \right] \ampd{\mu_\alpha  m_\gamma}{-\omega-i\varpi}
    = 
    - \mathbf{A}^{\mu_\alpha} t^{m_\gamma}
    - \mathbf{A}^{m_\gamma} \ampd{\mu_\alpha}{-\omega-i\varpi}
    - \mathbf{B} t^{m_\gamma} \ampd{\mu_\alpha}{-\omega-i\varpi}.
\end{multline}

However, the expression by which $\ampd{\mu_\alpha  m_\gamma}{-\omega-i\varpi}$ is multiplied is
exactly the right hand side of {the equations that determine} $\muld{\mu_\beta}{\omega+i\varpi}$, so {that} this term can be eliminated according
to:
\begin{multline}
\left[
         \mathbf{F} \ampd{\mu_\beta}{\omega+i\gamma} + \lhs{\mu_\beta}  
     \right] \ampd{\mu_\alpha m_\gamma}{-\omega-i\varpi}
     = 
     \\
     \muld{\mu_\beta}{\omega+i\gamma} \left[
         \mathbf{A}^{\mu_\alpha} t^{m_\gamma}
         + \mathbf{A}^{m_\gamma} \ampd{\mu_\alpha}{-\omega-i\varpi}
         + \mathbf{B} t^{m_\gamma} \ampd{\mu_\alpha}{-\omega-i\varpi}
    \right] .
\end{multline}
This leads to a more convenient computational expression,
which shows the symmetry between the perturbations:
\begin{multline}
    \label{damped_mcd_final}
    \frac{
        d \langle\langle \mu_\alpha; \mu_\beta \rangle\rangle_{\omega+i\varpi}
    }{
        d B_\gamma
    } = \;
    \mathbf{F}^{m_\gamma} \ampd{\mu_\beta}{\omega+i\varpi} \ampd{\mu_\alpha}{-\omega-i\varpi}
    \\+
    \Big[
        \mathbf{F}^{\mu_\alpha} \ampd{\mu_\beta}{\omega+i\varpi} +
        \mathbf{F}^{\mu_\beta} \ampd{\mu_\alpha}{-\omega-i\varpi}
    %\\
    +
        \mathbf{G} \ampd{\mu_\beta}{\omega+i\varpi} \ampd{\mu_\alpha}{-\omega-i\varpi}
    \Big] t^{m_\gamma}
    \\+
    \bar{t}^{m_\gamma} \Big[
        \mathbf{A}^{\mu_\alpha} \ampd{\mu_\beta}{\omega+i\varpi} +
        \mathbf{A}^{\mu_\beta} \ampd{\mu_\alpha}{-\omega-i\varpi}
    %\\
    +
        \mathbf{B} \ampd{\mu_\beta}{\omega+i\varpi} \ampd{\mu_\alpha}{-\omega-i\varpi}
    \Big]
    \\
    +
    \muld{\mu_\alpha}{-\omega-i\varpi}
    \big[
        \mathbf{A}^{\mu_\beta} t^{m_\gamma}
        +\mathbf{A}^{m_\gamma} \ampd{\mu_\beta}{\omega+i\varpi}
        %\\
        +\mathbf{B} t^{m_\gamma} \ampd{\mu_\beta}{\omega+i\varpi}
    \big]
     \\ 
     +
     \muld{\mu_\beta}{\omega+i\varpi}
     \big[
         \mathbf{A}^{\mu_\alpha} t^{m_\gamma}
         +\mathbf{A}^{m_\gamma} \ampd{\mu_\alpha}{-\omega-i\varpi}
         %\\
         +\mathbf{B} t^{m_\gamma} \ampd{\mu_\alpha}{-\omega-i\varpi}
     \big]
\end{multline}

{The connection to the quadratic response function expression
 $\langle\langle \mu_\alpha; \mu_\beta, m_\gamma\rangle\rangle_{\omega,0}$
(in the limit of $\varpi=0$)
is apparent.~\cite{MCD_CC,christiansen1998}}

\section{Results and discussion}
%\revS{Please fill in}
The calculation of the 
$\mathcal{A}_\mathrm{CC} (0 \rightarrow{f})$ and
$\mathcal{B}_\mathrm{CC} (0 \rightarrow{f})$ terms in gas phase according to the expressions in Eqs.~\eqref{MCD_Bterm_CC} and \eqref{MCD_Aterm_CC}
as well as that of the MCD ellipticity according to the CPP
algorithm discussed in Section~\ref{sec:damped} have been implemented at CCSD level in the 
our stand-alone python CC response platform.~\cite{FaberJCTC,rasmuspython} 
%\revS{Please fill in details about the geometries.}
%\rev{Please add details on how the stick spectrum data is broaded - put the formulas}
Two illustrative cases were considered,
cyclopropane, C$_3$H$_6$, and urea, H$_2$N(CO)NH$_2$.
Cyclopropane {has} D$_{3h}$ symmetry and thus possesses degenerate excited states, yielding 
spectral features
that arise 
from the ${\mathcal{A}}$-term.
Urea belongs to the C$_{2v}$ (or lower) point group and does not
support degenerate excited states per symmetry. 
% \revS{Comment on experimental spectra - we will be asked for sure!}
Experimental results in gas phase as well as computational (TDDFT and SOS-HF) 
results for cyclopropane are available in {the} literature.~\cite{Solheim2008,MCD_cyclopropane_Goldstein} To the best of our knowledge, the MCD spectrum of urea has neither been measured nor simulated before.

The geometry of urea was optimized at the CCSD/aug-cc-pVDZ level, whereas 
the geometry of cyclopropane was optimized at the CCSD(T)/aug-cc-pVTZ level.

The MCD spectra resulting from the 
calculated individual 
$\mathcal{A}_\mathrm{CC} (0 \rightarrow{f})$ and 
$\mathcal{B}_\mathrm{CC} (0 \rightarrow{f})$ terms were generated 
according to Eq.~\eqref{MCD_theta} with a Lorentzian lineshape function
\begin{align}
    g(\omega, \omega_f) =&
    \frac{\varpi}{\pi} \frac{1}{(\omega-\omega_f)^2+\varpi^2} \\
    \frac{\partial g(\omega, \omega_f)}{\partial \omega} =&
    -\frac{2\varpi}{\pi} 
    \frac{\omega-\omega_f}{\left[
        (\omega - \omega_f)^2 + \varpi^2
    \right]^2}
\end{align}
and the same $\varpi=0.0045563~\mathrm{a.u.} \approx 1000~\mathrm{cm}^{-1}$ was used for the broadening adopted in the damped response calculations.
The CCSD results are compared with CAM-B3LYP results obtained using LSDalton~\cite{DaltonPaper}. 
The values of the excitation energies and MCD terms for cyclopropane,
obtained from resonant response theory,
are collected in Table~\ref{tab:mcd_cyclopropane}.

\begin{table}[hbt!]
    \caption{
        \label{tab:mcd_cyclopropane}
        Computed spectral parameters for cyclopropane: excitation energies ($\omega_f$), dipole oscillator strengths (f),
        and MCD $\mathcal{A}$ and $\mathcal{B}$ terms.
    }
    \begin{ruledtabular}
    \begin{tabular}{llrr}
        %$\omega_f$ (a.u.) 
        Symm & $\omega_f$/eV (f) &  $\mathcal{A}$/a.u. & $\mathcal{B}$/a.u.\\
        \hline
        \multicolumn{4}{c}{CCSD/aug-cc-pVDZ}\\
        \hline
        %0.282470  
        E$^\prime$ & 7.686  (0.0001) &
        $-$0.00018785  
        &  0.24123551 \\   
    %    0.2984516688 &  0.0000000000 & -0.0000000000 \\   
    %    0.3030326616 (A1) &  0.0000000000 &  0.0000000000 \\   
    %    0.3039590926 &  0.0000000000 &  0.0000000000 \\   
        %0.305202 
        E$^\prime$
        & 8.305 (0.16)
        & 0.05915707 &  2.57266636 \\  
        %0.344009 
        E$^\prime$ & 9.361 (0.009)
        & $-$0.01141625 &  3.48280434 \\   
    %    0.3483933583 &  0.0000000000 &  0.0000000000 \\   
    %    0.3497780647 &  0.0000000000 & -0.0000000000 \\   
        %0.351224 
        A$^{\prime\prime}_2$ 
        & 9.557  (0.0098) &
        0.00000000 & $-$4.51643887 \\   
\hline
\multicolumn{4}{c}{CAM-B3LYP/aug-cc-pVDZ}\\
        \hline
        %0.274729  
        E$^\prime$
        & 7.476 (0.0001)
        & $-$0.00003014 &    0.18798269 \\  
%        0.28940278 &  0.00000000 &  0.00000000 \\   
%        0.29505860 &  0.00000000 &  0.00000000 \\   
%        0.29627551 &  0.00000000 &  0.00000000 \\   
        %0.297857  
        E$^\prime$
        &  8.105 (0.156)
        & 0.05278849 &    1.97266712 \\   
        %0.336915 
        E$^\prime$ & 9.168 (0.0088)
        & $-$0.01079584 &    2.96404589 \\   
%        0.33973464 &  0.00000000 &  0.00000000 \\   
%        0.34099048 &  0.00000000 &  0.00000000 \\   
        %0.341253 
        A$^{\prime\prime}_2$ &  9.286 (0.0096)
        & 0.00000000 & $-$3.61714300 \\   
    \end{tabular}
    \end{ruledtabular}
\end{table}

\begin{figure}[hbt!]
    %\centering
   % \includegraphics[width=\columnwidth]{cyclopropane_termsplit.pdf}
        \includegraphics[scale=0.7]{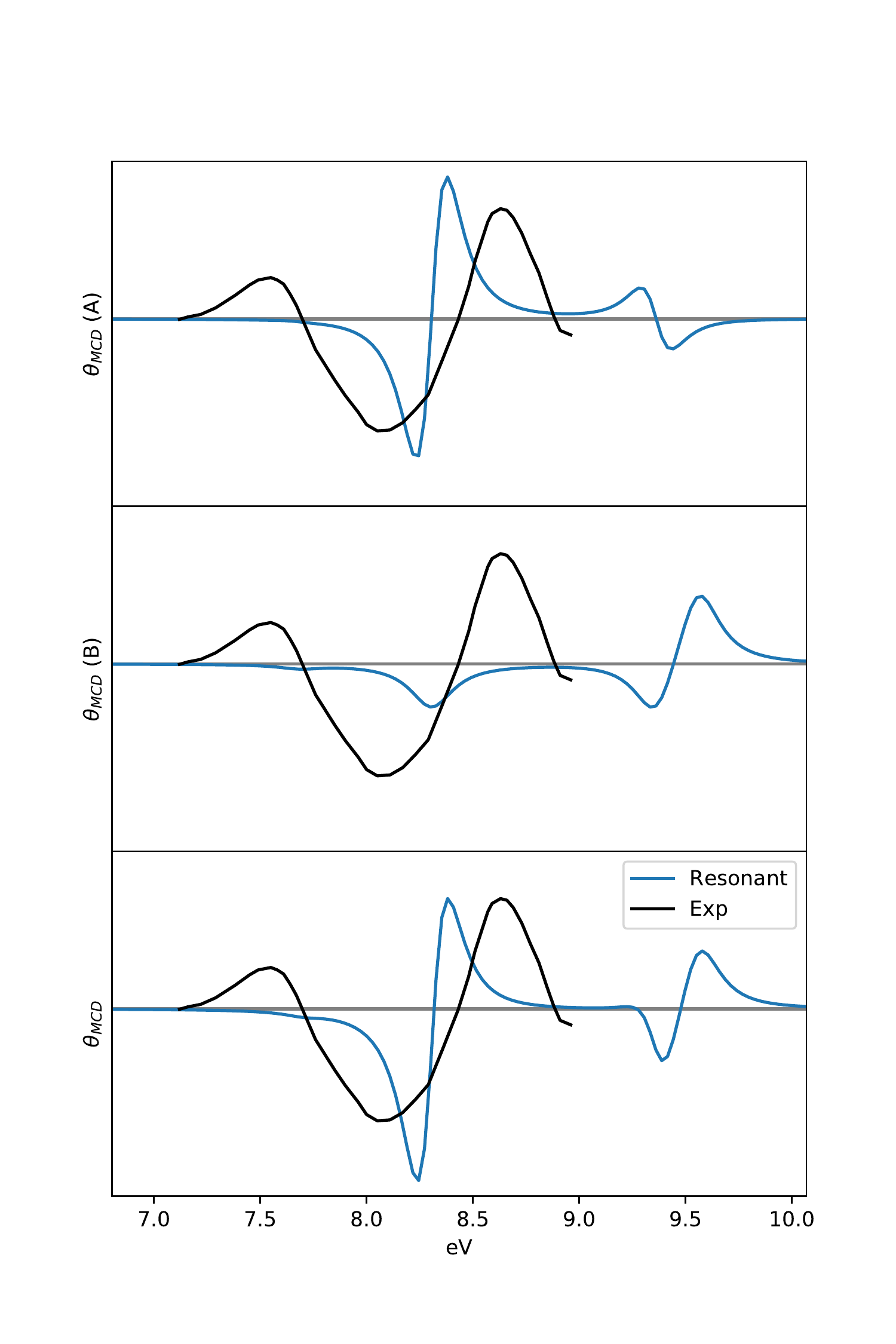}
    \caption{
 Cyclopropane. CCSD relative contributions of ${\mathcal{A}}_{\rm{CC}}$ (upper panel) and ${\mathcal{B}}_{\rm{CC}}$ (mid panel) terms of resonant response theory to the total (lower panel) broadened MCD spectrum. The experimental spectrum was taken from Ref.~\citenum{cyclopropane_exp}.
    }
    \label{fig:cyclopropane_termsplit}
\end{figure}

Based on the values in Table~\ref{tab:mcd_cyclopropane}, Figure~\ref{fig:cyclopropane_termsplit}
illustrates the relative importance of the ${\mathcal{A}}$ and ${\mathcal{B}}$ terms.
Clearly, the bisignate spectral feature centered at 8.30 eV is dominated by the positive 
${\mathcal{A}}$ term contribution of the second E$^\prime$ excited state, where  the ${\mathcal{B}}$ term is causing the slightly asymmetry of the dispersion band.
%A term is termed positive when DeltaA is positive to high energy
The second bisignate feature at around 
9.5~eV is the result of the fine balance of the negative ${\mathcal{A}}$ term for the third
E$^\prime$ state and the oppositely signed pseudo ${\mathcal{A}}$ due to the 
${\mathcal{B}}$ terms of the close-lying third
E$^\prime$ state and the non-degenerate A$^{\prime\prime}_2$ state.

\begin{figure}[hbt!]
    %\centering
    %\includegraphics[width=\columnwidth]{cyclopropane.pdf}
        \includegraphics[scale=0.6]{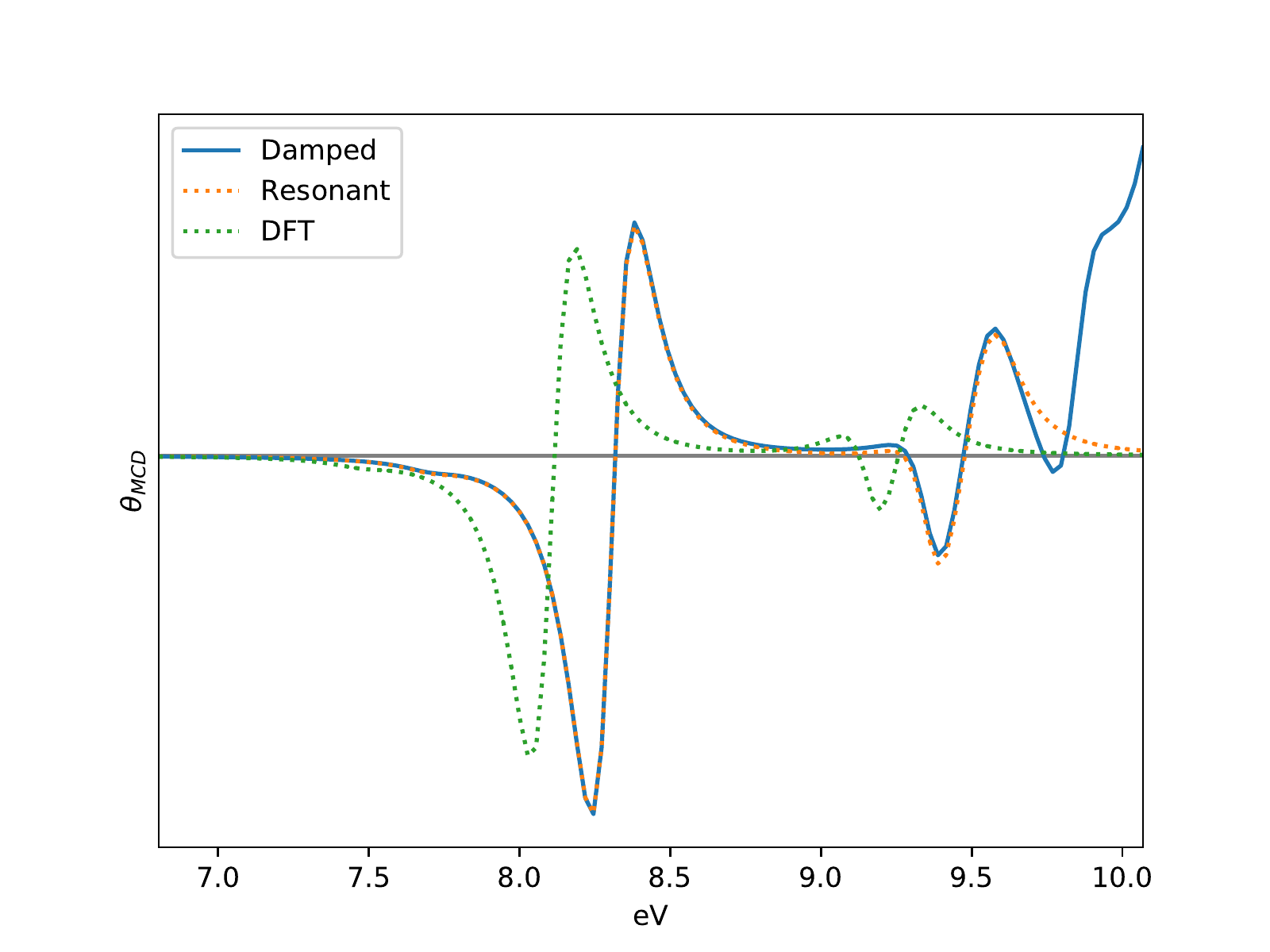}
    \caption{
        Cyclopropane. CCSD/aug-cc-pVDZ MCD spectra from damped and resonant response theory, and comparison with the CAM-B3LYP spectrum from resonant response theory.
    }
    \label{fig:cyclopropane_spec}
\end{figure}

The total MCD spectrum generated by Lorentzian broadening of the ${\mathcal{A}}$ and  
${\mathcal{B}}$ terms is compared with the spectrum obtained directly from damped response theory in Figure~\ref{fig:cyclopropane_spec}.
The broadened spectrum is basically identical to the one of damped response theory. The CAM-B3LYP spectrum is  red-shifted compared to the CCSD one, and with slightly weaker intensities, but otherwise 
the spectral profiles are  similar.

%\section{Results and discussion}

\begin{figure}[hbt!]
    %\centering
    \includegraphics[scale=0.7]{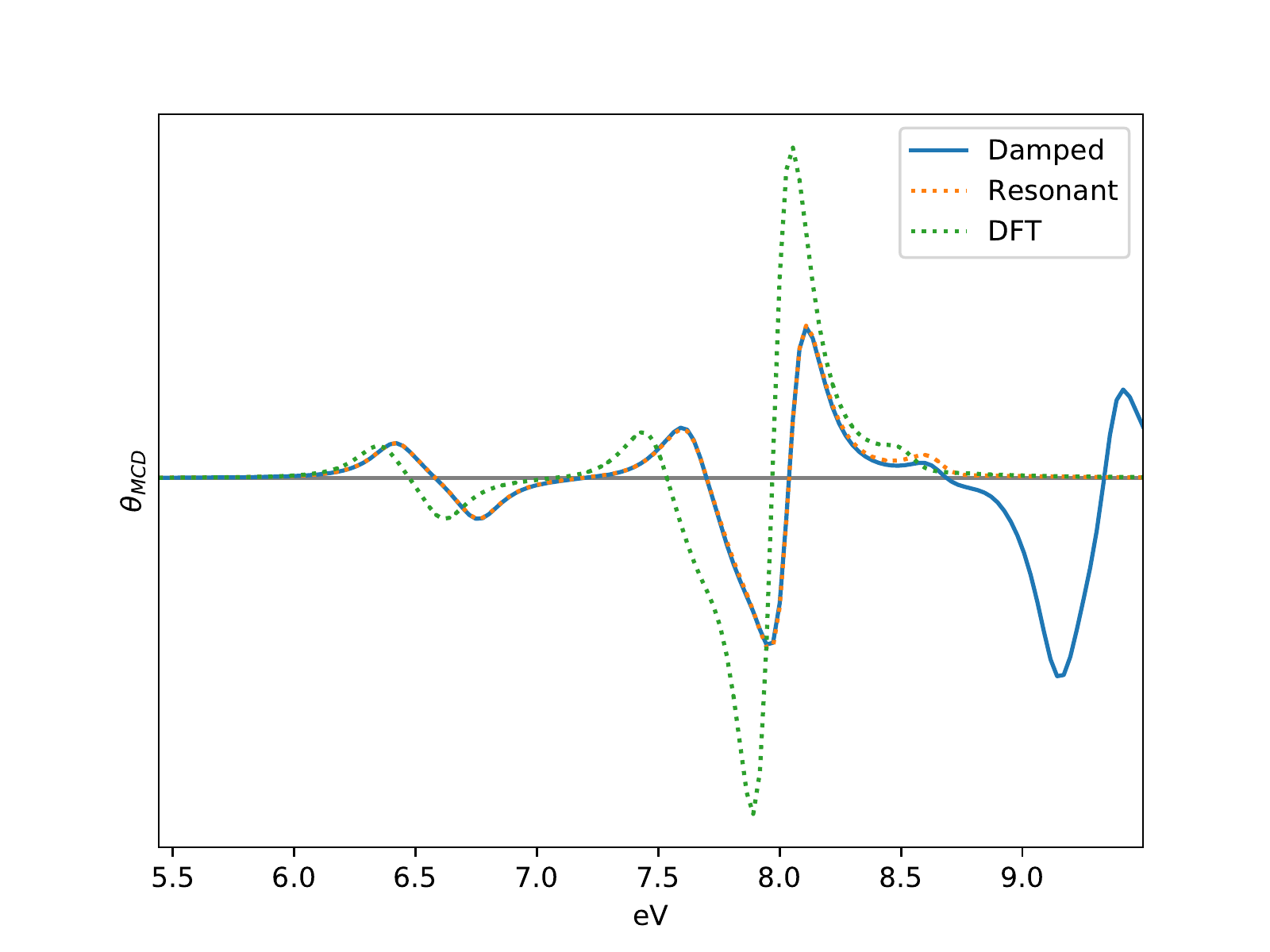}
    \caption{
        Urea. CCSD/aug-cc-pVDZ MCD spectrum from
        damped and resonant response theory, and comparison with the CAM-B3LYP one from resonant response theory.
    }
    \label{fig:urea_spec}
\end{figure}

Table~\ref{tab:mcd_urea} collects the values of the excitation energies and MCD spectral parameters for urea, as obtained from resonant response theory. 
Only ${\mathcal{B}}$ terms are possible by symmetry.
The corresponding MCD spectra, including the ones obtained from damped response, are shown in Figure~\ref{fig:urea_spec}.
Also in this case, damped and resonant theories yield almost identical spectra up to the number of frequencies that have been considered. The CAM-B3LYP spectrum is qualitatively very similar, yet with larger intensities from the two excited states at around 8 eV. 
%%%%%%%%% UREA %%%%%%

\begin{table}[hbt!]
    \caption{
        \label{tab:mcd_urea}
        Urea. Computed spectral parameters: excitation energies ($\omega_f$), oscillator strengths (f) and
        and MCD $\mathcal{B}$ 
        terms. Basis set aug-cc-pVDZ.}
    %\begin{ruledtabular}
    \begin{tabular}{lr|lr}
    \hline
\multicolumn{2}{c|}{CCSD}& 
\multicolumn{2}{c}{CAM-B3LYP}\\
        %$\omega_f$/a.u. 
        $\omega_f$/eV (f)&
        ${\mathcal{B}}$/a.u.&
        %$\omega_f$/a.u. 
        $\omega_f$/eV (f)&
        ${\mathcal{B}}$/a.u.\\
        \hline
%       0.2313096799 &  -0.0000000000 \\
        %0.23594 & 
        6.420 (0.031) & $-$4.78769 &
        %0.23352 & 
        6.354 (0.025) & $-$4.91873 \\ 
        %0.24820 & 
        6.754 (0.033) & 5.23536 &
        %0.24298 & 
        6.612 (0.033) & 5.66315 \\
%       0.2592198551 &   0.0000000000 \\
%       0.2747728769 &  -0.0000000000 \\
        %0.27649 & 
        7.524 (0.009) & $-$0.67620 &
        %0.27090 & 
        7.371 (0.012)& $-$0.62543 \\
        %0.28013 & 
        7.623 (0.036) & $-$7.46470 
        %& 0.27427 
        & 7.463 (0.021)& $-$7.63975 \\
        %0.28413 & 
        7.731 (0.014) &  1.53604  & 
        %0.27883 & 
        7.587 (0.020)&  1.83269 \\
        %0.28783 & 
        7.832 (0.002) &  5.05075 &
        %0.28201 & 
        7.674 (0.009)& 4.95339 \\
%       0.2906977582 &   0.0000000000 \\
        %0.29471 & 
        8.019 (0.14)& 84.0208  &
        %0.29137 & 
        7.928 (0.18)& 80.3061 \\
        %0.29597 & 
        8.054 (0.20) & $-$84.0705 &
        %0.29436 & 
        8.010 (0.13)& $-$80.2552 \\
%       0.3166729099 &  -0.0000000000 \\
        %0.31731 & 
        8.634 (0.059) & $-$4.23649 &
        %0.31304 & 
        8.518 (0.07)& $-$4.09754\\
        %0.31867 & 
        8.671 (0.003) &  3.05857 &
        %0.31446 &  
        8.557 (0.004)& 3.01059 \\\hline
    \end{tabular}
    %\end{ruledtabular}
\end{table}

\section{Conclusions}
We have presented a computational approach to obtain the ${\mathcal{A}}$ term of MCD within CC (resonant) response theory, together with alternative computational recipes for the ${\mathcal{B}}$ term,
Moreover, we have derived the  computational expression of the MCD ellipticity (temperature-independent part) within 
CC damped response theory. The latter can prove particularly convenient when the system under investigation is characterized by a large density of excited states.
Illustrative results have been reported for cyclopropane and urea, and compared with results from a previous B3LYP implementation of the ${\mathcal{A}}$ and ${\mathcal{B}}$ terms. The spectral profiles were found qualitatively similar, though with noticeable differences on the intensity scale
and the usual shifts in the
 position of the excited states.
%\clearpage

\begin{acknowledgments}
S.C. thanks Antonio Rizzo for useful discussion.
S.G. thanks the University of Brescia and MIUR for a visiting grant under the auspices of the Doctorate program.
R.F. and S.C. acknowledge financial support from the Independent Research
Fund Denmark – Natural Sciences, Research Project 2, grant no. 7014-00258B.
C.H. acknowledges financial support by the DFG through grant no. HA 2588/8.
\end{acknowledgments}

\section*{Data availability}
The data that support the findings of this study are available from the corresponding author upon 
reasonable request.

% \appendix

% \revS{
% \section{Appendix: alternative derivations/residue of the QRF}
% }

\section{Appendix: The residues of the derivative of the damped CC linear response function}
For the analysis of the residues of the derivative of $\langle\langle \mu_\alpha; \mu_\beta \rangle\rangle_{\omega+i\varpi}$ with respect to $B_\gamma$ we define the non-phase-isolated derivatives of the eigenvectors with respect to the electric fields:
\begin{align}
    R^{\mu_\alpha}_f(-\omega) & = -\big(\mathbf{A}-(\omega_f-\omega)\mathbf{1}\big)^{-1} \big(\mathbf{A}^{\mu_\alpha} + \mathbf{B} t^{\mu_\alpha}(\omega)\big) R_f
\end{align}
Of the amplitude and Lagrange multiplier vectors in Eq.~\eqref{eq.dCPPLR}
only the vectors $t^{\mu_\beta}(\omega)$, $\bar{t}^{\mu_\beta}(\omega)$, 
and $\bar{t}^{\mu_\alpha}(-\omega)$ have nonvanishing residues in the limit $\omega\to\omega_f$:
\begin{align}
    \lim_{\omega\to\omega_f} (\omega-\omega_f) t^{\mu_\beta}(\omega) & 
   % = \mathbf{R}_f \xi^{\mu_\beta} = R_f \cdot \big(L_f \xi^{\mu_\beta} \big) 
   = \sum_{f'\in \mathfrak{D}_f} R_{f'} T^{\mu_\beta}_{f'0}
\\
  \lim_{\omega\to\omega_f} (\omega-\omega_f) \bar{t}^{\mu_\beta}(\omega) & 
  % = \mathbf{F}\big(\mathbf{R}_f \xi^{\mu_\beta}\big)\mathbf{R}(-\omega_f) \\
  % & = \big(\mathbf{F}R_f \mathbf{R}(-\omega_f)\big) \cdot \big(L_f \xi^{\mu_\beta}\big) \\ 
   =  \sum_{f'\in \mathfrak{D}_f} T^{\mu_\beta}_{f'0} \cdot M_{f'}
\\
  \lim_{\omega\to\omega_f} (\omega-\omega_f) \bar{t}^{\mu_\alpha}(-\omega) & 
 % =  \Big[\eta^{\mu_\alpha} + \mathbf{F} t^{\mu_\alpha}(-\omega_f) \Big] \mathbf{R}_f
 % \\ & = \big(\eta^{\mu_\alpha} + M_f \xi^{\mu_\alpha} \big) \cdot L_f
  = \sum_{f'\in \mathfrak{D}_f} T^{\mu_\alpha}_{0f'} \cdot L_{f'}
\end{align}
We get for the non-singular part of residue (simple pole) for the limit $\omega\to\omega_f$:
\begin{align}
 & \lim_{\omega+i\varpi \to \omega_f}  (\omega+i\varpi-\omega_f)
       \frac{ d\rsp{\mu_\alpha}{\mu_\beta}_{\omega+i\varpi}}{d \epsilon_\gamma} \bigg|_{\text{non-res}}
\nonumber \\  
    =  & \frac{1}{2}\mathcal{P}^{\alpha\beta} \sum_{f'\in \mathfrak{D}_f} \bigg\{ \big(\mathbf{F}^{m_\gamma}  t^{\mu_\alpha}(-\omega_f) R_{f'}\big) \cdot T^{\mu_\beta}_{f'0}
 \\ \nonumber  &
 + \Big( \Big[\mathbf{F}^{\mu_\alpha} + \mathbf{G} t^{\mu_\alpha}(-\omega_f)  \Big] {R}_{f'} t^{m_\gamma}\Big) \cdot T^{\mu_\beta}_{f'0}
 \\ \nonumber & 
 + \Big( \bar{t}^{m_\gamma} \Big[\mathbf{A}^{\mu_\alpha} + \mathbf{B}  t^{\mu_\alpha}(-\omega_f) \Big] {R}_{f'} \Big) \cdot T^{\mu_\beta}_{f'0}
 \\ \nonumber &
 + \Big({^{\perp}\bar{t}}^{\mu_\alpha}(-\omega_f) \Big[\mathbf{A}^{m_\gamma} + \mathbf{B} t^{m_\gamma}  \Big] {R}_{f'} \Big) \cdot  T^{\mu_\beta}_{f'0}
 \\ \nonumber & 
 +
   L_{f'} \Big[\mathbf{A}^{\mu_\beta} t^{m_\gamma} + \mathbf{A}^{m_\gamma} {^{\perp}t}^{\mu_\beta}(\omega_f) + \mathbf{B} t^{m_\gamma} {^{\perp}t}^{\mu_\beta}(\omega_f) \Big] \cdot T^{\mu_\alpha}_{0f'} 
 \\ \nonumber &
 +
    M_{f'} \Big[\mathbf{A}^{\mu_\alpha}t^{m_\gamma} + \mathbf{A}^{m_\gamma} t^{\mu_\alpha}(-\omega_f) + \mathbf{B} t^{m_\gamma} t^{\mu_\alpha}(-\omega_f) \Big] \cdot T^{\mu_\beta}_{f'0}  \bigg\}
\\  
  = & \frac{1}{2}\mathcal{P}^{\alpha\beta} \sum_{f'\in\mathfrak{D}_f}  \bigg\{
  \Big( \bar{\xi}^{m_\gamma} R^{\mu_\alpha}_{f'}(-\omega_f) \Big) \cdot T^{\mu_\beta}_{f'0}
\\ & \nonumber
 +\Big( \bar{\xi}^{\mu_\alpha}(-\omega_f)  {^{\perp}R}_{f'}^{m_\gamma}\Big) \cdot T^{\mu_\beta}_{f'0} 
\\ & \nonumber
 + \Big( \Big[ \mathbf{F}^{m_\gamma} t^{\mu_\alpha}(-\omega_f)  + \mathbf{F}^{\mu_\alpha} t^{m_\gamma}  + \mathbf{G} t^{\mu_\alpha}(-\omega_f) t^{m_\gamma}\Big] R_{f'}\Big) \cdot T^{\mu_\beta}_{f'0} 
\\ & \nonumber
 + M_{f'} \Big[\mathbf{A}^{\mu_\alpha} t^{m_\gamma} + \mathbf{A}^{m_\gamma} t^{\mu_\alpha}(-\omega_f) + \mathbf{B} t^{m_\gamma} t^{\mu_\alpha}(-\omega_f) \Big]  \cdot T^{\mu_\beta}_{f'0} 
\\ & \nonumber
+ T^{\mu_\alpha}_{0f'} \cdot \Big( L_{f'} \mathbf{A}^{\mu_\beta} t^{m_\gamma} + {^{\perp}L}^{m_\gamma}_{f'} \xi^{\mu_\beta} \Big)
 \bigg\}
 \\
  = & \frac{1}{2}\mathcal{P}^{\alpha\beta} \sum_{f'\in\mathfrak{D}_f} \bigg\{
  \frac{d T^{\mu_\alpha}_{0f'}}{d\epsilon_\gamma} T^{\mu_\beta}_{f'0} + T^{\mu_\alpha}_{0f'} \cdot \frac{d T^{\mu_\beta}_{f'0}}{d\epsilon_\gamma}
 \bigg\}
 \end{align}
 Only one contribution to the derivative of the damped linear response function,
 $\bar{t}^{\mu_\alpha}(-\omega-i\varpi)\big[\mathbf{A}^{m_\gamma} + \mathbf{B} t^{m_\gamma}]t^{\mu_\beta}(\omega+i\varpi)$,
 %[{\bf{do you mean $\bar{t}^{\mu_\alpha}(-\omega)$}? ... yes.}]
 contains two vectors that become singular for $\omega\to\omega_f$ and contributes to the second-order residue:
 \begin{align}
 \lim_{\omega+i\varpi\to\omega_f} & (\omega+i\varpi-\omega_f)^2 \frac{d \rsp{\mu_\alpha}{\mu_\beta}_{\omega+i\varpi}}{d\epsilon_\gamma}
\\ & \nonumber
 = \frac{1}{2} \mathcal{P}^{\alpha\beta} \sum_{ {f,f'\in\mathfrak{D}_f}} 
%  \\ \nonumber &
   T^{\mu_\alpha}_{0f'} \cdot \Big(L_{f'} \Big[ \mathbf{A}^{m_\gamma}  + \mathbf{B} t^{m_\gamma} \Big] R_f \Big) T^{\mu_\beta}_{f0} 
\\ & =  \mathcal{P}^{\alpha\beta} \sum_{ f < f' \in\mathfrak{D}_f} 
  T^{\mu_\alpha}_{0f'}  \cdot T^{m_\gamma}_{f'f} \cdot T^{\mu_\beta}_{f0}
\end{align}
If we include the Levi-Civita tensor and the negative sign from Eq.~\eqref{MCD_damped_def},
the results agree with definition of the $\mathcal{B}$ and $\mathcal{A}$ terms in
Eqs.~\eqref{MCD_Bterm_CC} and~\eqref{MCD_Aterm_CC}.
%}

%\nocite{*}
%\bibliography{mcd}% Produces the bibliography via BibTeX.

\providecommand{\noopsort}[1]{}\providecommand{\singleletter}[1]{#1}%

\end{document}